\documentclass[
reprint,
floatfix,
amsmath,amssymb,aps,prb,superscriptaddress]{revtex4-2}

\usepackage{graphicx}
\usepackage{dcolumn}
\usepackage{bm}

\preprint{APS/123-QED}

\usepackage[separate-uncertainty=true]{siunitx}
\usepackage[version=4]{mhchem}
\usepackage{float}
\usepackage{mathtools}
\usepackage{hyperref}
\DeclareSIUnit\samples{\text{S}}
\usepackage[no-test-for-array]{nicematrix}
\usepackage{nicefrac}

\begin{document}
    \title{Laser-Induced Relaxation Oscillations in Superconducting Nanobridge Single Photon Detectors}
    
    \author{F.B. Baalbergen}
    \email{baalbergen@physics.leidenuniv.nl}
    \affiliation{Huygens-Kamerlingh Onnes Laboratory, Leiden University, P.O. Box 9504, 2300RA Leiden, The Netherlands}
    \author{I.E. Zadeh}
    \affiliation{Department of Imaging Physics (ImPhys), Faculty of Applied Sciences, Delft University of Technology, 2628CJ Delft, The Netherlands}
    \author{M.J.A. de Dood}
     \email{dood@physics.leidenuniv.nl}
    \affiliation{Huygens-Kamerlingh Onnes Laboratory, Leiden University, P.O. Box 9504, 2300RA Leiden, The Netherlands}
    
    \date{\today}
    
    \begin{abstract}
        We demonstrate novel laser-induced relaxation oscillations in superconducting nanowire single photon detectors (SNSPDs).
        These oscillations appear when a voltage biased \ce{NbTiN} nanobridge detector is illuminated with intense pulsed laser light at a repetition rate of $\sim\qty{19}{\MHz}$.
        They differ from the well-known relaxation oscillations by a step-wise increase in frequency and phase locking of the oscillations to the laser pulses.
        We create a model that incorporates electrical feedback and excludes thermal effects to simulate and explain the origin of the observed laser-induced relaxation oscillations.
        Qualitative agreement to the experiment is achieved using realistic values for the parameters in the model.
    \end{abstract}
    
    \maketitle
    \section{Introduction}
    Superconducting Nanowire Single Photon Detectors (SNSPDs) are a successful technology for applications in quantum optics and can enable optical quantum computation.
Using SNSPDs provides a great advantage over other technologies due to the fast recovery times, low timing jitter, high detection probabilities~\cite{Chang2021,Reddy2020,Hu2020} and possibly native photon number resolution~\cite{Zhu2020, Los2024}.
For quantum information applications, such as quantum key distribution or quantum state preparation, high detection rates with high detection probability are needed\cite{Gruenenfelder2023}. 
A better understanding of electrical-thermal feedback in SNSPDs proved to be an essential step in speeding up photon detection~\cite{Liu2012, Liu2013, Yang2007, Annunziata20102, Kerman2009}.

Over the years, different biasing methods were presented to provide a bias current to an SNSPD.\
These methods include a quasi-constant current source using a resistor in series with a voltage source~\cite{Kerman2006,Annunziata20102}, a quasi-constant voltage source using a shunt resistor in order to limit latching of the detector into the resistive state~\cite{Liu2012, Liu2013} as well as various ways of cryogenic readout and biasing~\cite{Kitaygorsky2009,Kerman2013,Thiele2024}.
In this work we choose to use a quasi-constant voltage bias with values of the bias resistors chosen to prevent latching of the detector.
Previous work using a similar biasing circuit shows clear relaxation oscillations due to overbias when the detector is not illuminated with laser light~\cite{Liu2012, Liu2013}.
In this work we show that the biasing electronics enables laser-induced relaxation oscillations, therefore eliminating other possible mechanisms, such as after pulsing~\cite{Raupach2023}.

In this work we present the first observations of laser-induced relaxation oscillations in nanoscale bridges of \qtyrange{70}{150}{\nano\metre} width.
We present a simple electrical model that can be used to model the laser-induced relaxation oscillations and achieve qualitative agreement to the experimental data. 
We analyze and interpret these synchronized oscillations in terms of electrical feedback mechanisms in SNSPD's.
The observed relaxation oscillations limit the maximal count rate for high bias conditions to rates that are well below the inverse of the detector reset time.
This becomes particularly important for infrared photon detection where SNSPDs are biased close to the critical current and for high count rates where electrical feedback reduces the effective bias current.
Here we show the effect for \qty{780}{\nm} light for bias currents well below $I_\text{SW}$.
Observing the laser induced relaxation oscillations then requires higher laser powers.
A better and more complete understanding of the feedback mechanisms responsible for laser induced relaxation oscillations in SNSPDs may thus help to achieve faster count rates in future devices by better understanding the limitations of devices imposed by the biasing circuit.
    
    \section{Methods}
    Nanobridge SNSPDs are fabricated out of \qty{13}{\nm} thin film \ce{NbTiN} deposited on a \ce{Si (100)} substrate with a thermal \ce{SiO2} layer of \qty{230}{\nm} thickness and the device is capped with a \qty{12}{\nm} thick \ce{Si3N4} layer.
The nanobridge SNSPDs are defined using standard e-beam lithography and etching techniques~\cite{Zichi2019} and consist of a nanowire constricted to a section of equal length and width of \qtylist{70;100;120;150}{\nano\metre}.
To prevent latching, the nanobridges are connected in series with a \qty{500}{\nm} wide meandering wire with $\sim\qty{700}{\nano\henry}$ inductance to slow down the response of the detector.
The \qty{13}{\nm} thickness of the \ce{NbTiN} film results in detectors that all have the same measured relatively high critical temperature $T_c=\qty{9.1(1)}{\kelvin}$ with a slope of \qty{6}{\mega\ohm\per\kelvin} at the superconducting transition (see appendix \ref{sec:app:cooldown}), where we define the critical temperature as the temperature where the resistance is half of the maximum measured resistance ($R_\text{max}=\qty{4.34}{\mega\ohm}$). 

\begin{figure}
    \centering
    \includegraphics[width=85mm]{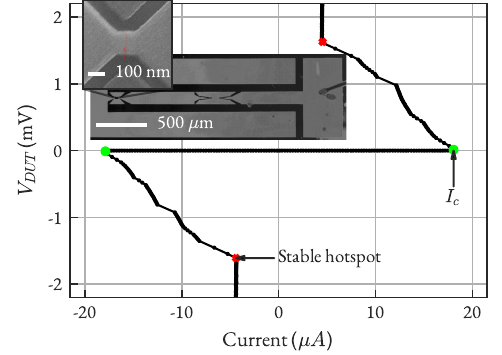}
    \caption{Measured I-V curve of a \qtyproduct{120 x 120}{\nano\meter} \ce{NbTiN} nanobridge SNSPD at $T=\qty{6}{\kelvin}$ under voltage bias.
    The critical current of the nanobridge is $\sim\qty{18}{\micro\ampere}$ (green dots).
    A stable hotspot appears for voltages $|V_{DUT}|>\qty{1.5}{\milli\volt}$ (red crosses).
    Oscillations are observed in the regime between the critical current and the stable hotspot.
    The inset shows an optical image of the entire structure of contacts, inductor and nanodetector together with an SEM image of the nanodetector.}\label{fig:IV}
\end{figure}

\begin{figure}
    \centering
    \includegraphics[width=85mm]{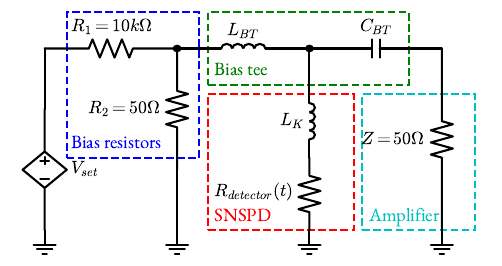}
    \caption{Schematic of the electrical circuit with the SNSPD represented by an equivalent circuit of a kinetic inductance $L_K$ in series with a time-dependent resistance $R_\text{detector}(t)$.
    The SNSPD is connected via a bias-tee (mini-circuits parts ZNBT-60-1W+ or ZFBT-6GW+).
    A voltage bias is applied at the DC port of the bias-tee using a voltage source (Yokogawa GS200) and a set of resistors $R_1$ and $R_2$ to create a quasi-current source.
    The RF port of the bias-tee is connected to room temperature amplifiers (mini-circuit ZFL-500LN+) with an input impedance $Z = \qty{50}{\ohm}$.}\label{fig:circuit}
\end{figure}

Figure~\ref{fig:IV} shows the detector structure (see inset) together with the measured I\nobreakdash-V curve for a \qty{120}{\nm}\nobreakdash-wide nanobridge SNSPD at a temperature of \qty{6}{\kelvin}.
The I\nobreakdash-V curve is measured with a 2\nobreakdash-wire setup and biased using a quasi-constant voltage bias in order to prevent latching~\cite{Liu2012}.
The detector is connected using the electrical circuit shown in Fig.~\ref{fig:circuit}.
A voltage source (Yokogawa GS200) is used in combination with $R_1=\qty{10}{\kilo\ohm}$ and $R_2=\qty{50}{\ohm}$ resistors.
The value of $R_2$ is close to the maximum value that avoids latching.
In this configuration relaxation oscillations start as soon as there is a slight overbias of the detector.
The detector is connected to a bias tee (minicircuits ZNBT\nobreakdash-60\nobreakdash-1W+) and a \qty{50}{\ohm} AC-coupled amplifier ($2\times$ minicircuits ZFL\nobreakdash-500LN+), both at room temperature. 
The SNSPD is mounted in a closed-cycle cryostat (Entropy GmbH) and is represented by the equivalent circuit consisting of a kinetic inductance $L_K$ and a time dependent resistance $R_{\text{detector}}(t)$.
The measurements presented in this work are all done at a temperature of \qty{6}{\kelvin}.

The current through the detector is determined by monitoring the voltage over resistor $R_2=\qty{50.9(5)}{\ohm}$ using a digital multimeter (Keithley 2000).
The voltage over the detector is the corrected voltage measured over $R_2$.
This voltage is corrected for the resistance of the semi-rigid cables ($\sim$7$\Omega$) and a thermal offset by defining $V_\text{set}=\qty{0}{\volt}$ at the point where the current $I_\text{detector}=\qty{0}{\micro\ampere}$. 

To measure the detector response to light, the detector is illuminated by a pulsed picosecond supercontinuum laser with a $\sim\qty{19}{\MHz}$ master oscillator repetition rate.
The output of the laser is filtered using a \qty{1000}{\nm} shortpass filter and a \qty{780}{\nm} bandpass filter with a \qty{10}{\nm} FWHM bandwidth.

The detector is mounted in a cryostat with free-space optical access and the laser is focused on the detector.
The average power is varied between \qty{0.5}{\nano\watt} and \qty{50}{\uW} using a motorized stage with a $\frac{\lambda}{2}$\nobreakdash-plate between two crossed Glan\nobreakdash-Thompson prisms (B. Halle) that acts as polarizers and achieve $\sim$50 dB attenuation.

The output pulses from the detector after the amplifiers are either recorded using a digital counter (Agilent 53220A) or a digitizer at a sample rate of \qty{5}{\giga\samples\per\second} and an analog bandwidth of \qty{3}{\giga\hertz} (Teledyne ADQ7).

    \section{Results}
    The measured I\nobreakdash-V curves in Fig. 1 show that the \qty{120}{\nm} detector has a critical current of \qty{18.1(1)}{\micro\ampere} at $T=\qty{6}{\kelvin}$ (green circles in figure~\ref{fig:IV}).
Taking into account the cross-sectional area of the nanobridge this corresponds to a critical current density of \qty{1.2\pm0.1e6}{\ampere\per\centi\meter^2}.
Measurements on nanobridges of other dimensions show similar critical current densities.

Oscillations occur under voltage bias once the set voltage is increased beyond the point where the critical current is reached~\cite{Kerman2009}.
At this point the time-averaged voltage over the device under test (DUT) $V_\text{DUT}$ becomes non-zero while the time-averaged current through the device decreases when the voltage is increased.
This regime persists until a stable, self-heating, hotspot is formed in the nanobridge (red crosses in figure~\ref{fig:IV}).
For the \qty{120}{\nm} wide detector we find a stable hotspot resistance of \qty{370(10)}{\ohm} at \qty{4.5}{\micro\ampere}.
The square resistance of the superconducting film is estimated to be $\sim\qty{500}{\ohm}$ as calculated from the resistance of the on chip inductor.

\begin{figure*}
    \centering
    \includegraphics[width=179mm]{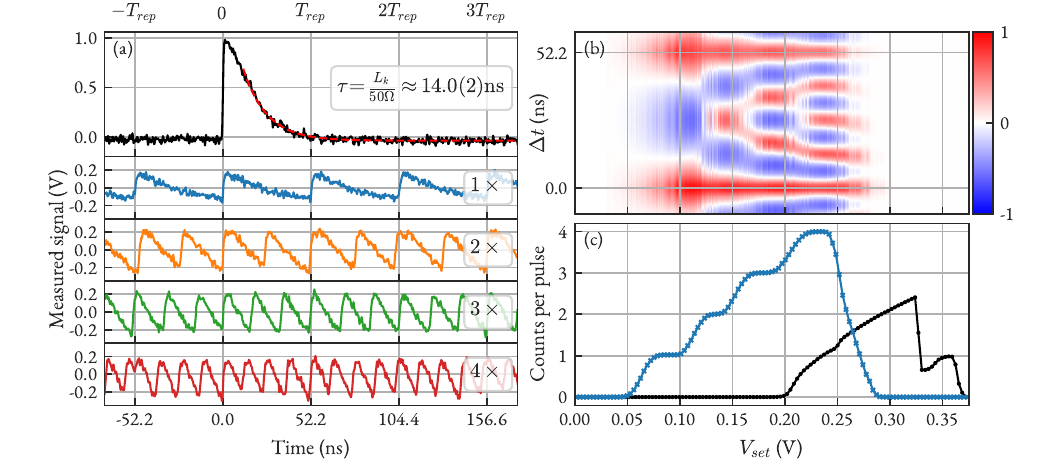}
    \caption{Observation of laser-induced relaxation oscillations.
    \textbf{(a)} Time traces measured at the output of the amplifiers. 
    From top to bottom pulses are shown for a dark event, a train of pulses at the laser repetition rate of $\sim\qty{19}{\mega\hertz}$ and $2$, $3$ and $4$ times the repetition rate. 
    The dashed line for the dark event is a fit to an exponential decay with $\tau = \qty{14.0\pm 0.2}{\nano\second}$ ($L_K \approx \qty{700}{\nano\henry}$). 
    \textbf{(b)}Autocorrelation of the detector pulses as a function of bias voltage and
    \textbf{(c)} measured count rate normalized by the laser repetition rate as a function of bias voltage for a device in the dark (black crosses) and under intense pulsed laser illumination (blue circles, \qty{63.30(7)}{\micro\watt}). 
    The lines through the data serve to guide the eye and show clear stepwise transitions in the oscillation rate up to 4 times the laser repetition rate. 
    The first step of the illuminated device is caused by the detector becoming an efficient, saturated detector, the light induced relaxation oscillations start with the second step at $V_\text{set}=\qty{0.1}{\volt}$.
    In the data for the device in the dark, three distinct regimes can be identified, these regimes correspond to the non-smooth behavior of the detector in Fig.~\ref{fig:IV}}\label{fig:results}
\end{figure*}

Figure~\ref{fig:results} shows the measured laser-induced relaxation oscillations for the \qty{120}{\nm} nanobridge.
Time traces of the detector response are shown in Fig.~\ref{fig:results}a and show, from top to bottom, a dark event (black trace) and relaxations oscillations under intense pulsed laser illumination.
Using the measured laser power of \qty{10}{\micro\watt}, the laser spot size, the geometric area of the detector, and the expected absorption of the nanowire we estimate that 5 to 10 photons are absorbed by the detector per laser pulse(see appendix \ref{sec:app:photon_response}).
The laser power is kept constant in our experiment while the voltage over the device is increased. 
With increasing set voltage we observe novel `quantized' relaxation oscillations that occur at integer multiples of the laser repetition rate of \qty{19}{\MHz} (1$\times$, 2$\times$, 3$\times$ and 4$\times$ labels).

We analyze the dark event in Fig.~\ref{fig:results}a to estimate the reset time.
From the fit to an exponential decay we find a reset time $\tau = \nicefrac{L_k}{R} = \qty{14.0\pm0.2}{\nano\second}$ and estimate a kinetic inductance of $ L_k\approx\qty{700\pm10}{\nano\henry}$~\cite{Kerman2006}.

Examples of time traces at different values of $V_{set}$ are shown in Fig.~\ref{fig:results}a.
The figure shows oscillations at integer multiples of the laser repetition rate.
The long term stability and phase locking behavior of these time traces become obvious by computing the auto-correlation defined as
\begin{align}
R_{ff}(\tau)&=\int_\mathbb{R}f(t+\tau)f(t)\mathrm{d}t\bigg/\int_\mathbb{R}f(t)^2\mathrm{d}t
\end{align}
from the recorded digitizer traces. 
This integral is discretized and taken over the complete length of the data trace.\footnote{The calculation is performed using the \texttt{scipy.signal.correlate} function~\cite{Virtanen2020}.}
Each time trace has a total length of \qty{200}{\us} giving a total uncompressed size of \qty{800}{\mega\byte} for the $400$ data traces used for Fig.~\ref{fig:results}b.

Fig.~\ref{fig:results}b shows the autocorrelation of the laser-driven oscillations.
The autocorrelation is a measure for the self similarity of a signal at different time scales $\tau$.
Clear phase-locking behavior can be seen as the autocorrelation has large peaks at integer fractions of time between two laser pulses.
The extra counts occur exactly between two light pulses and show the even distribution expected for a phase-locked oscillation.

To further explore the novel relaxation oscillations, we vary the bias voltage and measure the count rate as a function of set voltage for the detector under intense illumination and without illumination.
Fig.~\ref{fig:results}c shows the results of these experiments, where the count rates are normalized to the $\sim\qty{19}{\MHz}$ repetition rate of the laser.
The figure shows normalized count rates as a function of $V_{set}$ both in the light (blue curve) and the dark (black curve).

Under intense laser illumination the count rate first increases from $0$ to $1$ counts per pulse around $V_{set} = \qty{0.06}{\volt}$.
A careful inspection of the time traces in this regime reveals that this onset of detection is caused by the amplitude of the detection peaks becoming larger than the discriminator level in the counter.
The discriminator level is set just above the electronic noise level of the amplifier. 
We define the point $V_{set} = \qty{0.10}{\volt}$ as the point beyond which clear laser-induced relaxation oscillations occur.
The frequency of these oscillations increases with both bias voltage and optical power.
For higher optical powers we see that the point where the laser induced relaxation oscillations occur, shifts to lower set voltages.
Interestingly, the relaxation oscillations in this regime phase-lock to the laser pulses and the oscillation frequency becomes an integer multiple (up to $4\times$ for the \qty{120}{\nm} detector) of the $\sim\qty{19}{\MHz}$ laser repetition rate, shown as plateaus in Fig.~\ref{fig:results}c.
We note that the bias voltage where clear laser-induced relaxation oscillations start is much lower than the bias voltage of $V_{set} = \qty{0.20} {\volt}$ where relaxation oscillations start when the detector is not illuminated.

The novel laser-induced relaxation oscillations are well-defined oscillations and the mechanism differs from the free relaxation oscillations where the oscillation frequency increases continuously with increasing set voltage~\cite{Liu2012, Liu2013, Kerman2009}.

    \section{Model}
    In the following section we introduce a model to simulate the laser-induced relaxation oscillations.

In the model the electrical part of the SNSPD is represented by an equivalent circuit.
This circuit is shown in Fig.~\ref{fig:circuit} and consists of a time-dependent resistor ($R_\text{detector}$) in series with an inductor ($L_k$). 
This resistance $R_\text{detector} = \qty{0}{\ohm}$ when the detector is in the superconducting state.
In the model we assume that the time-dependent resistor switches instantaneously between the superconducting state and the normal state of the detector where $R_\text{detector}=R_{HS}=\qty{3}{\kilo\ohm}$.
In our experiment the inductor is the kinetic inductance of a nanofabricated long meandering wire in series with the nanobridge that avoids latching of the detector after a detection event~\cite{Kerman2009, Yin2024, Annunziata20102}.
Based on the dimensions of the nanobridge and a value of the sheet inductance of the \ce{NbTiN} we estimate that the kinetic inductance of the nanobridge is below \qty{0.05}{\nano\henry} and can be neglected compared to the \qty{700}{\nano\henry} of the series inductor.

To simplify the simulation of the finite thermal response of the nanobridge, we assume that the detector in our model stays in the normal resistive state for a fixed time of $t_0=\qty{0.5}{\nano\second}$.
The values for the hotspot resistance and response time are based on previously published values for the hotspot dynamics after photon absorption~\cite{Yang2007, Berggren2018, Kong2022, Zhao2018}. 
To include the measured resistance of the coaxial cable in the cryostat we set $R_\text{detector}=\qty{7}{\ohm}$ in the superconducting state.
Introducing this resistance is important because the value is not negligible compared to the \qty{50}{\ohm} resistance in the bias circuit and the \qty{50}{\ohm} input impedance of the amplifier.
Retardation effects are not included in our circuit model and hence the \qty{50}{\ohm} impedance of the coaxial cable should be ignored.

A hotspot is formed in the model whenever a laser pulse arrives and when the current exceeds the device critical current $I_c$.
All the dynamics of the detector are included in the time and current dependence of $R_\text{detector}$.
This simplified model, that ignores thermal transport, is sufficient to mimic the behavior of the nanobridge including relaxation oscillations. 
In the model we use the measured value of $I_c = \qty{18}{\micro\ampere}$. 
The response to pulsed laser light is added to the simulations through a probability $p$ that switches the detector when the detector is illuminated. 
These events occur at times that are defined by the $\qty{19}{\MHz}$ repetition rate of the laser. 
The intense laser pulses in our experiment correspond to a situation where multiple photons are absorbed per pulse, corresponding to $p = 1$.

The output of our model is defined as the voltage over the $Z=\qty{50}{\ohm}$ impedance of the amplifier on the AC-port of the bias-tee. 
The model is solved using circuit theory by repeatedly solving for the voltages and currents in the system (see Appendix). 
The approximate values for the inductance and capacitance in the bias-tee can be estimated by using the cutoff frequencies $\omega=1/(RC)$ and $\omega=R/L$ of the bias-tee. 
Here we take the $Z=\qty{50}{\ohm}$ line impedance for the resistance and the specified cutoff frequency from the manufacturer. 
We have repeated our experiments with different bias-tees (Minicircuits ZNBT\nobreakdash-60\nobreakdash-1W+ and ZFBT\nobreakdash-6GW+) to confirm that the choice of bias-tee does not significantly affect our results. 
Simulations from the model confirm that the observation of the laser-induced relaxations oscillations do not depend on the values of the capacitance and inductance in the bias-tee. 
Hence, the only free parameters in our simulation are the hotspot resistance $R_{HS}$ and the minimal hotspot time $t_0$. 

\begin{figure}
    \centering
    \includegraphics[width=85mm]{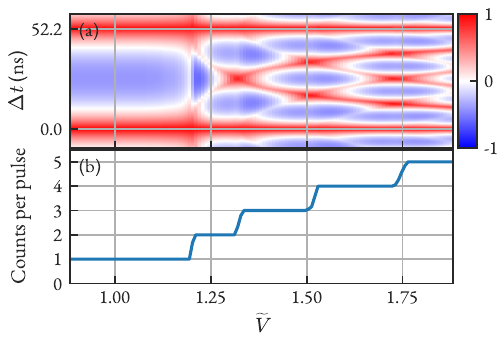}
    \caption{Simulated laser-induced relaxation oscillations, based on the electronic model depicted in Fig.~\ref{fig:circuit}. 
    \textbf{(a)} Auto-correlation of the simulated time traces showing the synchronization at integer multiples of the laser repetition rate.
    \textbf{(b)} Count rate as a function of the unitless detector bias $\widetilde{V}$. 
    Simulation parameters are set to realistic values (see text) to resemble the experimental observations.
    Clear steps are observed, demonstrating the synchronization of relaxation oscillations at integer values of the laser repetition rate.}\label{fig:correlation_model}
\end{figure}

The results for this simulation can be found in Fig.~\ref{fig:correlation_model} and shows qualitative agreement with our experimental observations.
Here we simulate the system for different realistic values of $V_{set}$.
Calculations for $V_{set}$ between \qtyrange{0}{0.35}{\volt}, $I_c$ from \qtyrange{10}{20}{\micro\ampere} and $R_1$ from \qtyrange{5}{15}{\kilo\ohm} show that all results converge to a universal curve 
\begin{align}
    R\left(\frac{V_{set}}{I_c R_1}\right)=R(\widetilde{V})
\end{align}
where $R$ is the measured count rate and we define a unitless parameter $\widetilde{V}$.
Despite the fact that our model does not contain any details about heat generation, heat diffusion or timescales, we find laser-induced relaxation oscillations that resemble the experimental result.
Fig.~\ref{fig:correlation_model}a shows the calculated auto-correlation using the model output as a function of $\widetilde{V}$.
It should be noted that the non-constant step size in the model originates from the exponential recovery of the bias current after detection.

    \section{Discussion}
\begin{figure}
    \centering
    \includegraphics[width=85mm]{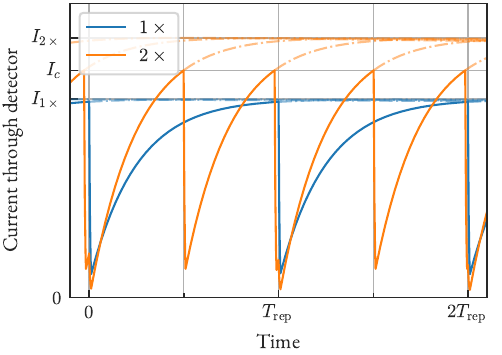}
    \caption{
        Current through the detector during laser-induced relaxation oscillations as predicted by the model.
        Curves shown for a count rate at a rate equal to the laser repetition rate (blue curve) and laser-induced relaxation oscillations at two times the laser repetition rate (orange curve).
        The dashed lines indicate the exponential recovery of the detector to the level $I_{1\times}$ and $I_{2\times}$ (see text).
        }\label{fig:discussion}
\end{figure}

    The intuitive explanation of the observed laser-driven relaxation oscillations is as follows:
In our electrical model, the hotspot resistance and minimal hotspot time determine the average resistance of the detector.
The average resistance of the detector feeds back into the bias circuit through the inductance of the bias tee.
Therefore the voltage over the resistor $R_2$ will be higher for a given $V_{set}$ compared to the system in the dark.
This causes the current in the recovery phase after photon detection to recover to a higher current compared to the dark.
This is a well-known phenomenon in AC coupled readout in SNSPDs using a current source~\cite{Kerman2013}.
Whenever this average resistance is high enough to push the current beyond the device critical current, a hotspot forms which feeds back into the average resistance.
As a result the average current through the detector reduces due to the higher average resistance of the detector.
This feedback mechanism causes the laser-induced relaxation oscillations to synchronize to (multiples of) the laser repetition rate. 
In Fig.~\ref{fig:discussion} a schematic representation of this phenomenon is shown.
When increasing $V_{set}$, it becomes possible for this process to find a different stable point causing higher order oscillations.

The described model is a powerful tool in understanding the laser-induced relaxation oscillations. 
The model reproduces all features found in the experiment.
Besides this, there are minor limitations to the model.

We emphasize that the described model does not contain a thermal description of the SNSPD device nor of the substrate.
The agreement with the experimental data rules out an electro-thermal feedback mechanism as a primary cause for the laser induced relaxation oscillations.
A consequence of the simplified model is that the model does not contain the required physics to describe latching of the detector.
Consequently, our model predicts that the frequency of relaxation oscillations grows without bound.
In the experiment, the laser-induced relaxation oscillations stop and a stable hotspot is formed.
Once a stable hotspot is created a detector will stay in this hotspot position. 
This mechanism has been explored before to explain and limit latching behavior of detectors due to relaxation oscillations due to overbias of a detector without illumination~\cite{Liu2012}. 
In our simplified model the physics of the stable hotspot is not included and therefore the number of steps in the laser-induced relaxation oscillations and maximum oscillation frequency are not bounded.

A complete description would also link the laser-induced relaxation oscillations to the geometry of the detectors.
In the experiment we find that wider wires show lower maximum frequencies of laser-induced relaxation oscillations.
This can be explained by the higher critical current due to the wider wire.
This causes the balance between the stable hotspot formation and the laser-induced relaxation to shift more towards stable hotspot formation.

Finally, the model assumes a constant critical current independent on the amount of light present.
Due to the small detection area in the detectors most of the light on the detector is absorbed in the area around the active detection area.
Therefore we suspect that the local temperature of the detector is higher compared to the rest of the sample.
This causes the critical current of the device to be reduced because the higher local temperature.
Due to this lowering of $I_c$, the laser-induced relaxation oscillations will be shifted to lower bias points.

The difference in step size in the model is not as pronounced in the experimental data. 
We attribute this to heating effects that decrease $I_c$ and hence decrease the step size.
The formation of a stable hotspot in the experiment further limits the amount of observable steps.

    \section{Conclusion}
    In conclusion, we demonstrate the first observation of laser synchronized relaxation oscillations in nanobridge SNSPDs at high photon flux.
These oscillations occur when using a quasi constant voltage bias while illuminating the detector with intense pulsed laser light.
Unlike the commonly observed dark relaxation oscillations, these laser-induced relaxation oscillations are synchronized and phase-locked to the laser pulses.
A simplified electrical model of the SNSPD that ignores the thermal properties is sufficient to simulate the laser-induced relaxation oscillations with reasonable values for the hotspot resistance and duration of the normal state.
Synchronization to the laser pulses leading to oscillations at rates that are an integer multiple of the laser repetition rate are explained as an electronic feedback between the time average resistance of the detector and the bias current through the detector.
This feedback mechanism changes the biasing condition of the detector as a function of the average count rate (and thus average resistance) of the detector and thus enables stable operating points at frequencies that are an integer multiple of the repetition rate of the laser.

    \begin{acknowledgments}
    We thank M.P. van Exter for helpful discussions and J.C. Rivera Hernández and K. Kanneworff for setting up the first measurements on the samples. 
    This research was made possible by financial support of the Dutch Research Council (NWO).
    \end{acknowledgments}

    \appendix
    \begin{widetext}
    
    \section{Simulation of laser-induced relaxation oscillations}\label{sec:app:simulation}
    \subsection{Time-dependent nodal analysis}
The properties of superconducting nanowires and bias-circuits can be modeled using SPICE~\cite{Berggren2018} provided that a sufficient number of parameters is given to correctly model the dynamical behavior of the hot-spot or normal region.
Our simplified model reduces this dynamics by assuming a non-latching detector with a constant high-resistance in the normal state for a preset duration.
As a result, our simplified model ignores thermal feedback effects.
The current-dependent switching of the nanowire is included in the model by checking for overbias, i.e.\ detector bias exceeding a switching current, at every timestep.

To solve the output for the electronic circuit we make use of a time-dependent nodal analysis that introduces a set of equations for the current and voltage at the nodes in the electronic circuits.
These equations introduce the current and voltage as a column vector $\vec{V}_i$ and $\vec{i}_i$.
The state of the circuit is described by a state vector $\vec{S}$ that is a column vector defined by $\vec{S} = (\vec{V}_i, \vec{i}_i)^T$.

The goal of the nodal analysis is to find update equations and solve for the time-dependent output. 
We split the time steps for the current and voltage by defining $\vec{S}_{-} = (\vec{V}_i(t-\Delta t), \vec{i}_i (t-\Delta t/2))^T$, $\vec{S}_{0} = (\vec{V}_i(t), \vec{i}_i (t-\Delta t/2))^T$ and $\vec{S}_{+} = (\vec{V}_i(t), \vec{i}_i (t+\Delta t/2))^T$. 
We can define the update operations $\mathbf{A}$, $\mathbf{B}$ and $\mathbf{U}=\mathbf{B}\cdot\mathbf{A}$ as the matrix such that $\mathbf{A}:\vec{S}_-\mapsto\vec{S}_0$, $\mathbf{B}:\vec{S}_0\mapsto\vec{S}_+$ and $\mathbf{U}:\vec{S}_-\mapsto\vec{S}_+$.
These matrix equations calculate the next state of the system given that the time step $\Delta t$ is sufficiently small.

\subsection{Circuit equations}
\begin{figure}[H]
    \centering
    \includegraphics[width=85mm]{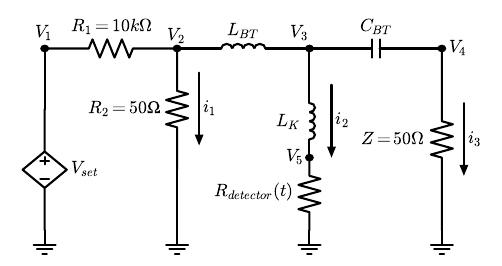}
    \caption{Schematic of the electrical circuit in the simulation indicating the 5 independent voltage nodes $V_1 \ldots V_5$ and the 3 independent current paths $i_1 \ldots i_3$.}\label{fig:nodal}
\end{figure}
For the specific circuit of the SNSPD we identify 5 independent voltage nodes and 3 currents paths as shown in Fig.~\ref{fig:nodal}.
The equations for the nodes and currents follow from the Kirchhoff current and voltage laws.

\begin{align}
    V_1(t)-V_2(t) &= R_1\left(i_1(t)+i_2(t)+i_3(t)\right)\\
    V_2(t) &= R_2 i_1(t)\\
    V_3(t)-V_2(t) &= -L_{bt}\frac{\mathrm{d}(i_2(t) + i_3(t))}{\mathrm{d}t}\\
    V_5(t)-V_3(t) &= -L_k\frac{\mathrm{d}i_2(t)}{\mathrm{d}t}\\
    V_3(t)-V_4(t) &= \frac{1}{C_{BT}}\int_0^t i_3(\tau)\mathrm{d}\tau\\
    V_4(t) &= Z i_3(t)\\
    V_5(t) &= R_{\det}(t,i_2) i_2(t)
\end{align}

\subsection{Discrete circuit equations}
We solve these equations numerically and introduce discrete time steps $\Delta t$ that we denote using square brackets for reasons of clarity.
In principle the circuit can be analyzed using SPICE as an open source analog circuit simulator~\cite{Nagel1973}.
Unfortunately, the implementation of the time-dependent resistor that mimics the behavior of the SNSPD in the equivalent circuit by solving a one-dimensional heat transport problem in combination with switching events due to overbias is cumbersome \cite{Berggren2018}.
The full simulation, introduces additional parameters that need to be determined independently and slows down the simulations.
We circumvent these issues by creating a dedicated numerical solution for our specific circuit where we check the state of the system at each time step.
The detailed thermal response of the nanobridge is then reduced to a value of the normal state resistance and a timescale.

We write the time-discretized voltages as
\begin{align}
    V_2[t] & = R_{2}\cdot i_1\left[t-\tfrac{1}{2}\Delta t\right]\\
    V_5[t]& = R_{\det}(t,i_2)\cdot i_2\left[t-\tfrac{1}{2}\Delta t\right]\\
    V_4[t]& = Z\cdot i_3\left[t-\tfrac{1}{2}\Delta t\right]\\
    V_3[t]-V_4[t]& = V_3\left[t-\tfrac{1}{2}\Delta t\right]-V_4\left[t-\tfrac{1}{2}\Delta t\right]+ \frac{\Delta t}{C_{BT}}\cdot i_3\left[t-\tfrac{1}{2}\Delta t\right]
\end{align}

Using the updated voltages we find the currents using
\begin{align}
    V_5[t]-V_3[t] &= -\frac{L_k}{\Delta t}\left(i_2\left[t+\tfrac{1}{2}\Delta t\right] - i_2\left[t-\tfrac{1}{2}\Delta t\right]\right)\\
    V_3[t]-V_2[t] &= -\frac{L_{BT}}{\Delta t}\left(i_2\left[t+\tfrac{1}{2}\Delta t\right] - i_2\left[t-\tfrac{1}{2}\Delta t\right]+i_3\left[t+\tfrac{1}{2}\Delta t\right] - i_3\left[t-\tfrac{1}{2}\Delta t\right]\right)\\
    V_1[t]-V_2[t] &= R_1(i_1\left[t+\tfrac{1}{2}\Delta t\right]+i_2\left[t+\tfrac{1}{2}\Delta t\right] +i_3\left[t+\tfrac{1}{2}\Delta t\right])
\end{align}

Rearranging these equations yields the following for the currents

\begin{align}
    i_2\left[t+\tfrac{1}{2}\Delta t\right] &= i_2\left[t-\tfrac{1}{2}\Delta t\right]-\frac{\Delta t}{L_k}\left(V_5[t]-V_3[t]\right)\\
    i_3\left[t+\tfrac{1}{2}\Delta t\right] &= i_3\left[t-\tfrac{1}{2}\Delta t\right]-\frac{\Delta t}{L_{BT}}\left(V_3[t]-V_2[t]\right)+\frac{\Delta t}{L_k}\left(V_5[t]-V_3[t]\right)\\
    i_1\left[t+\tfrac{1}{2}\Delta t\right] &= \frac{V_1[t]-V_2[t]}{R_1}-i_2\left[t+\tfrac{1}{2}\Delta t\right]-i_3\left[t+\tfrac{1}{2}\Delta t\right]
\end{align}

\subsection{Update matrices}
From these equations we can find the update matrices $\mathbf{A}$ and $\mathbf{B}$

\begin{align}
    \mathbf{A} &=
    \begin{pNiceArray}{ccccc|ccc}
        1&0&0&0&0&0&0&0\\
        0&0&0&0&0&R_2&0&0\\
        0&0&1&-1&0&0&0&Z+\frac{\Delta t}{C_{BT}}\\
        0&0&0&0&0&0&0&Z\\
        0&0&0&0&0&0&R_{\det}(t,i_2)&0\\
        \hline
        \Block{3-5}<\Large>{\mathbf{0}} & & & & & \Block{3-3}<\Large>{\mathbb{I}_3}& & \\
        & & & & & & &\\
        & & & & & & &\\
    \end{pNiceArray}
\intertext{and}
    \mathbf{B} &=
    \begin{pNiceArray}{ccccc|ccc}
        \Block{5-5}<\Large>{\mathbb{I}_5} & & & & & \Block{5-3}<\Large>{\mathbf{0}} & & \\
        & & & & & & &\\
        & & & & & & &\\
        & & & & & & &\\
        & & & & & & &\\
        \hline
        \frac{1}{R_1}&-\left(\frac{1}{R_1}+\frac{\Delta t}{L_{BT}}\right)&\frac{\Delta t}{L_{BT}} &0&0&0&-1&-1\\
        0&0&\frac{\Delta t}{L_k}&0&-\frac{\Delta t}{L_k}&0&1 &0\\
        0&\frac{\Delta t}{L_{BT}}&-\left(\frac{\Delta t}{L_{BT}}+\frac{\Delta t}{L_K}\right)&0&\frac{\Delta t}{L_k}&0&0 &1
    \end{pNiceArray}
\end{align}

where $\mathbb{I}_3$ and $\mathbb{I}_5$ denote the $3\times3$ and $5\times5$ unit matrices, respectively. 
The structure of these matrices clearly show the consecutive update of the voltages and currents.

The total update matrix $\mathbf{U} = \mathbf{B}\cdot\mathbf{A}$ is used in the numerical code to update the state of the system.
We achieve good convergence of the numerical method using a time step $\Delta t = \qty{20}{\pico\second}$ for representative values of $L_{BT} = \qty{200}{\micro\henry}$, $C_{BT} = \qty{30}{\nano\farad}$, $L_k = \qty{700}{\nano\henry}$, $R_{HS} = \qty{3}{\kilo\ohm}$ and $R_\text{cable} = \qty{7}{\ohm}$.

    \end{widetext}
    
    \section{Single photon response}\label{sec:app:photon_response}
    
\begin{figure}[H]
    \centering
    \includegraphics[width=85mm]{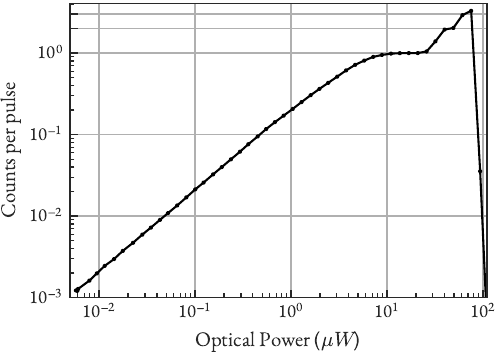}
    \caption{Measured count rate of a \qtyproduct{120 x 120}{\nano\meter} \ce{NbTiN} nanobridge SNSPD as a function of optical power. Measurements are performed at $T=\qty{6}{\kelvin}$ using $\sim\qty{19}{\mega\hertz}$ pulsed $\qty{780}{\nano\meter}$ laser light.
    The linear power dependence of the count rate as a function of power confirms that the nanodetector responds to single photons.
    The single photon detector saturates for powers above \qty{10}{\micro\watt} and is pushed into laser induced relaxation oscillations and the stable hotspot regime at maximum optical power.}\label{fig:light_response}
\end{figure}

    Single photon detectors show a linear increase in count rate as a function of optical power.
Figure~\ref{fig:light_response} shows the measured count rate of the detector as a function of the average optical power for a detector biased below the critical current.
 
As can be seen, the slope of the data on a log-log scale is linear at low powers and saturates at a click probability of $1$ for optical powers of $\sim\qty{10}{\micro\watt}$.
For the highest optical powers the detector enters a regime of laser induced relaxation oscillations as a function of optical power ending in a stable hotspot regime with no counts at the highest power.
Both the bias voltage and the click probability are constant in this regime  and we attribute these oscillations to heating effects from the laser which reduce $I_c$ and lead to overbias of the detector at high optical powers.

A more detailed look at the data extrapolates the linear power dependence to predicts a saturation power of \num{3}--\qty{5}{\micro\watt}.
Given the laser repetition rate we estimate that saturation sets in at \num{0.15}--\qty{0.25}{\pico\joule\per\text{pulse}}, corresponding to \num{4}--\qty{6e5}{\text{photons}\per\text{pulse}}.
Assuming a close to $100\%$ internal quantum efficiency of the detector saturation occurs at one absorbed photon per pulse and we estimate the efficiency of the detector to be \numrange[range-phrase = --]{0.1e-6}{0.2e-6}.
A similar value of the efficiency can be estimated from the absorption of the thin \ce{NbTiN} film on the substrate ($\sim25\%$) and the ratio of the detector area and the diffraction limited spot size (beamwaste of approximately \qty{10}{\micro\meter}) in our setup. 

Depending on the exact value of the saturation power we estimate that the average number of absorbed photons in the regime where laser-induced relaxation oscillations are observed is 5--10 photons per pulse. 

\begin{figure}[H]
    \centering
    \includegraphics[width=85mm]{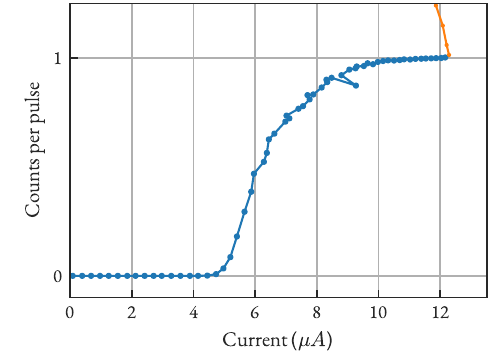}
    \caption{Click probability of a \qtyproduct{120 x 120}{\nano\meter} \ce{NbTiN} nanobridge SNSPD as a function of bias current. 
    Measurements are performed at $T=\qty{6}{\kelvin}$ using $\sim\qty{19}{\mega\hertz}$ pulsed $\qty{780}{\nano\meter}$ laser light at an average power of \qty{13.6(2)}{\micro\watt} ($\sim 1$ photon absorbed per pulse).
    The blue curve (large dots) shows the detector response before entering the relaxation oscillation regime.
    The orange curve (small dots) show the onset of laser induced relaxation oscillations.}\label{fig:sigmoidal}
\end{figure}

The detector response as a function of bias current at low power settings ($\sim 1$ photon absorbed per pulse) is shown in figure~\ref{fig:sigmoidal}.
A saturation is clearly visible with near $100\%$ click probability for $I>\qty{10}{\micro\ampere}$ ($I/I_{sw}~0.6$).

    \section{I-V curves of illuminated detector}\label{sec:app:IV}
    
\begin{figure}
    \centering
    \includegraphics[width=85mm]{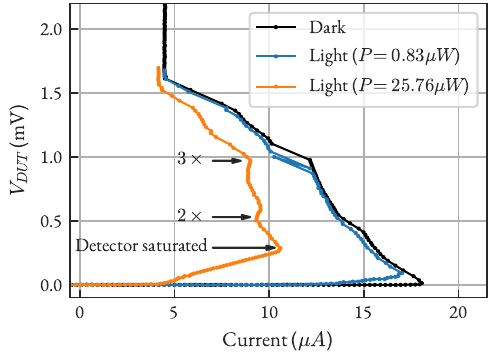}
    \caption{Measured I-V curves of a \qtyproduct{120x120}{\nano\meter} \ce{NbTiN} nanobridge SNSPD at $T=\qty{6}{\kelvin}$ illuminated with different amount of laser light.
    Data are shown for a detector without illumination (black curve), in the single photon regime (blue curve) and in the saturated regime where laser-induced oscillation occur (orange curve).
    The arrows in the graph indicate the onset of different regimes for the orange curve.}\label{fig:IV_light}
\end{figure}

    In figure~\ref{fig:IV_light} the IV-curve for the detector is shown for three photon fluxes.
The black line shows the same data as presented in figure~\ref{fig:IV}.
The blue and orange curves are measured with pulsed laser illumination of the detector.
For the blue curve, the detector is far from the regime of laser induced relaxation oscillations.
Photon counting events are observed and the effect of these events is seen in the high current regime of the superconducting phase.
In this regime the average measured voltage over the detector is higher due to the photon counting events.
The maximum average current of the detector is decreased due to counting.
The orange curve is measured with a much more optical power at the detector.
Due to this optical power, the detector has a much higher (almost unity) chance of producing a counting event for a light pulse.
Therefor, the average voltage over the detector starts going up for lower bias currents as compared to the blue curve.
Furthermore, at this light level, we also see laser induced relaxation oscillations.
For the orange curve, these locations are shown by arrows in the figure.

    \section{Superconducting transition R(T)}\label{sec:app:cooldown}
    
\begin{figure}
    \centering
    \includegraphics[width=85mm]{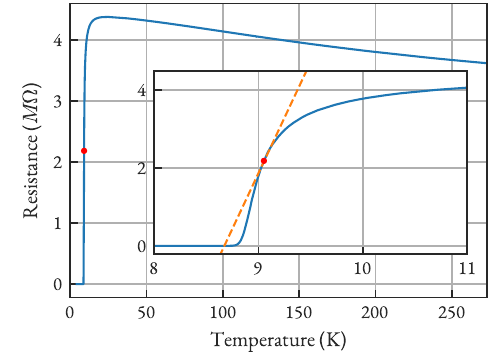}
    \caption{Measured 2 wire resistance as a function of the temperature for a \qtyproduct{70 x 70}{\nano\meter} detector using a \qty{700}{\nano\ampere} probe current.
    The inset shows the data around $T_C$.}\label{fig:cooldown}
\end{figure}

    Figure~\ref{fig:cooldown} shows the measured resistance of a \qtyproduct{70x70}{\nm} detector on the same sample during cooldown.
Measurements are performed using a 2-wire configuration with a \qty{700}{\nano\ampere} probe current from a digital multimeter.
The data shows a maximum resistance of \qty{4.34}{\mega\ohm} at $T=\qty{24}{\kelvin}$ and a room temperature resistance of \qty{3.57}{\mega\ohm}.

The (approximate) transition is found at a temperature $T_c=\qty{9.1(1)}{\kelvin}$ as indicated by the circle in the inset of the figure.
The straight line through the data around $T_c$ indicate the slope $\mathrm{d}R/\mathrm{d}T$ at the transition.

\end{document}